\documentclass[conference]{IEEEtran}

\ifCLASSINFOpdf

\else
  
\fi

\usepackage{graphicx}
\usepackage{amsmath,mathtools}
\usepackage{amssymb}
\usepackage{amsthm}
\usepackage[shortlabels]{enumitem}
\usepackage{caption}
\usepackage{subcaption}
\usepackage{dcolumn}
\usepackage{bm}
\usepackage{mathrsfs}
\usepackage{xcolor}
\usepackage{color}
\usepackage{float}
\usepackage{xcolor}
\usepackage[normalem]{ulem}
\usepackage{bbold} 
\usepackage{cite}

\usepackage{url}
\usepackage{comment}

\begin{document}
%
\title{Optimal Navigation on Simplicial Complexes}

\author{
\begin{tabular}{cc}
\begin{minipage}[t]{0.45\textwidth}
\centering
\textbf{Diego Febbe}\\
Department of Physics and Astronomy\\
University of Florence, Sesto Fiorentino, Italy\\
INFN Section of Florence, Sesto Fiorentino, Italy\\
Email: diego.febbe@unifi.it
\end{minipage}
&
\begin{minipage}[t]{0.45\textwidth}
\centering
\textbf{Duccio Fanelli}\\
Department of Physics and Astronomy\\
University of Florence, Sesto Fiorentino, Italy\\
INFN Section of Florence, Sesto Fiorentino, Italy\\
Email: duccio.fanelli@unifi.it
\end{minipage}
\\[2.3cm]
\begin{minipage}[t]{0.45\textwidth}
\centering
\textbf{Gianluca Peri}\\
Department of Physics and Astronomy\\
University of Florence, Sesto Fiorentino, Italy\\
INFN Section of Florence, Sesto Fiorentino, Italy\\
Email: gianluca.peri@unifi.it
\end{minipage}
&
\begin{minipage}[t]{0.45\textwidth}
\centering
\textbf{Timoteo Carletti}\\
Department of Mathematics\\
University of Namur, Namur, Belgium\\
Institute for Complex Systems -- naXys\\
Email: timoteo.carletti@unamur.be
\end{minipage}
\end{tabular}
}

\maketitle

\begin{abstract}
The navigation time and optimal search strategies deriving from random dynamical processes on binary graphs have been extensively 
explored and analyzed, being of prominent interest in the network science field.
In this work, we study an extension of these topological measures for simplicial complexes: a specific type of geometric and algebraic 
structures that encapsulates higher-order interactions. 

Here, the \textit{explorability} analysis of simplicial complexes has been conducted in terms of the mean first passage times between nodes, 
i.e. the $0^\textsuperscript{th}$ $\!\!\!\!$-order simplices, with the inclusion of a long-range
  stochastic teleportation term modulated with respect to the local random walk hopping across the various dimensions. 
We also provide a perturbative approximation scheme recovering the modulation parameter between pure random walk and teleportation mode (for higher order setting) acting as the expansion parameter. 

\end{abstract}

\IEEEpeerreviewmaketitle

\section{\label{sec:Introduction} Introduction\protect
}
\begin{figure}[h]
    \centering
    \includegraphics[width=0.99\linewidth]{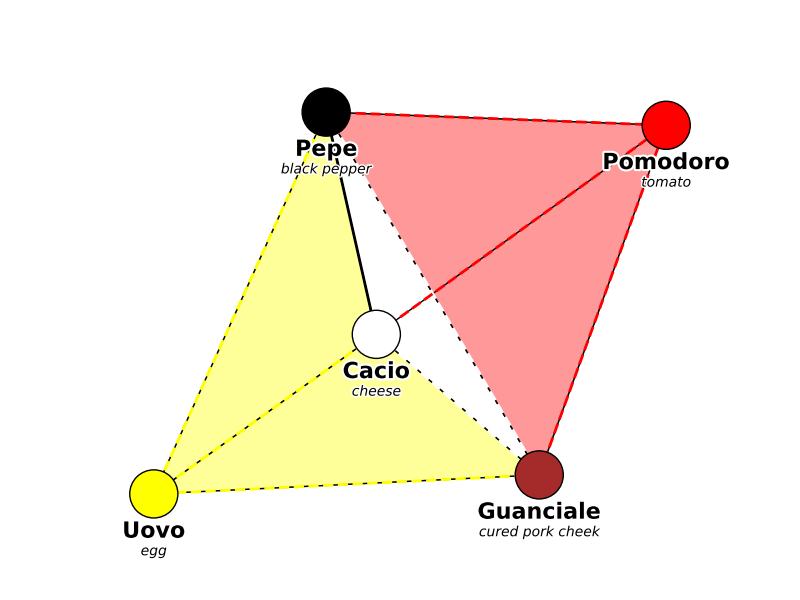}
    \caption{Example of intuitive visual representation of a higher-order network of ingredients that compose famous Italian pasta sauce recipes. 
    The solid black link connects pepper (black node) and cheese (white) forming the \textit{cacio e pepe} (other two-body interactions are represented as dashed lines since they do not corresponding to famous combinations).  The two former ingredients, 
    in combination with cured pork cheek (brown), can form the three-body interaction named \textit{gricia} (white triangle). 
    Then, by adding either egg (yellow) or tomato (red) you can form the celebrated \textit{carbonara} or \textit{amatriciana} (yellow and red tetrahedra).
    Despite the lack of physical formalism, this may be a very easy and concrete way to introduce higher-order interactions.}
    \label{fig:food_ho}
    \vspace{-5mm}
\end{figure}

Network theory has proven to be of extreme importance for the modeling of interacting complex systems \cite{barabasi2016network, newman2010networks}
with a plethora of uses, spanning from biological and neuro-scientific studies \cite{zhang2005general, barabasi2023neuroscience, lucas2023inferring},
 engineering \cite{nishikawa2015comparative, febbe2024chaos} and artificial intelligence applications 
 \cite{corso2024graph, buffoni2022spectral, chicchi2026estimating}, the formulation of socio-economic 
 problems or even the description of epidemic spreading and opinion formation in social contexts \cite{pacelli2025link, munoz2025exploring, centola2010spread, buongiovanni2022will}.
Recently, however, some portion of the scientific community has started to expand the description of interactions 
in order to include the multi-body interplay (beyond the binary case) into the modeling of complex systems \cite{bianconi2021higher, battiston2020networks}. 
Examples can be found in a variety of scientific fields: from theoretical physics \cite{bianconi2025gravity, lucarini2026synergistic}, 
to social dynamics \cite{iacopini2019simplicial}, neural network architectures \cite{morris2019weisfeiler, peri2026spectral, peri2026spectralhigherorderneuralnetworks} or neuroscience 
\cite{giusti2016two}. Some intuitive and pictorial examples of real systems showing 
higher-order interactions can be found in the network representation of co-authorship collaborations 
for scientific publications \cite{Benson-2018-simplicial, vasilyeva2021multilayer} or of ingredients required for cooking 
recipes \cite{fujisawa2022analyzing} (see Fig. \ref{fig:food_ho}).

Higher-order networks, and in particular simplicial complexes, offer also a prolific ground for the study of dynamical processes 
involving stability studies, synchronization and Turing pattern formation 
\cite{gambuzza2021stability,  gallo2022synchronization, carletti2023global, giambagli2022diffusion, muolo2023turing, muolo2024turing, moriame2025hamiltonian, leon2024higher}.
Simplices, the fundamental elements composing simplicial complexes, are a particular type of hypergraphs enriched with the so-called \textit{closure property}, 
an algebraic condition that includes all the sub-interactions (links, triangles, ... ) made of nodes present in them. 
Various algorithms for simplicial complexes construction are presented and extensively analyzed in the works 
\cite{bianconi2016network, torres2020simplicial, bianconi2017emergent,  febbe2026model} while the study of the reciprocal interplay between their topology and the dynamical processes unfolding on them is of timely interest
 \cite{millan2025topology, febbe2026random}. 
Examples of cross-dimensional dynamics among the various simplicial complex orders, have been invoked in biological models and in particular, neuroscientific systems \cite{giusti2016two, petri2014homological} where neurons are
connected via synapses and are influenced by higher-order interactions stemming from neural cliques of co-firing elements. 
A conceptually basic, yet fundamental, description of cross-dimensional dynamics on simplicial complexes is presented in \cite{febbe2026random} 
where a random walk process occurring between the various orders is formulated in terms of the normalized-unsigned version of the complete 
\textit{Dirac operator} naturally encapsulating the structural algebraic properties \cite{bianconi2021topological}. This yields a clear connection between the network topology, describing multi-body interactions, and the dynamical properties. 
The basic version of this latter random walk process is defined by allowing one-order jumps among simplices of various dimensions, 
but in a direct generalization, the local dynamics can be combined with global stochastic teleportation. 
This is the foundational idea of the celebrated \textit{PageRank} algorithm \cite{page1999pagerank} and optimal search strategies 
for its variants are analyzed in \cite{di2015optimal} and generalized for simplicial complexes in \cite{febbe2026random}.
Here, we build upon this by presenting further analysis on the higher-order explorability, 
in terms of the mean first passage time in the presence of long-range jumps. Moreover, we provide a perturbative scheme to recover the optimal searching configuration explicitly depending on the modulating parameter.
This work is organized as follows: in Sec. \ref{sec:Model} we will briefly summarize the random walk formulation on simplicial complexes. 
Then, in Sec. \ref{sec:OSS_approximation} we will present the explorability and optimal search strategy analysis and finally in 
Sec. \ref{sec:Conclusion} we will draw the main conclusions.
The code for the simplicial complexes generation and the random walk process implementation is present in \cite{diegofebbe_2026_21563618} and maintained on the GitHub repository\footnote{\url{https://github.com/diegofebbe/Random_walk_on_simplicial_complexes/tree/master}}.

\section{\label{sec:Model} Random Walk on Simplicial Complexes}

\begin{figure}[h]
    \centering
    \includegraphics[width=0.8\linewidth]{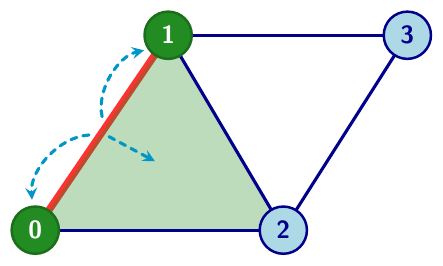}
    \caption{Pictorial representation of the random walk process described in \cite{febbe2026random}. 
    Let us start with a walker initialized on the simplex $\sigma^{(1)}=[0,1]$ at discrete time $t$ (bold red line in figure). 
    On the next time step it can jump to one among all the one-dimension upper- and lower-connected simplices, namely the nodes 
    [0] and [1] and the triangle [0,1,2] (green in figure).}
    \label{fig:rw_sc}
\end{figure}

Recently, in \cite{febbe2026random} a random walk process on simplicial complexes has been formulated in terms of the unsigned boundary operators 
of the various orders, inducing a dynamical process that naturally travels across structures that are connected on the corresponding boundaries 
(nodes with links, links with nodes and triangles etc...). Fig. \ref{fig:rw_sc} offers an illustrative one-discrete-time-step representation 
of the above-described process.
More formally, by retracing the main steps of this random walk formulation \cite{febbe2026random}, let us consider a $D$-simplicial complex, 
$\mathcal{X}$, containing $N_d$ $d$-simplices, $0\leq d \leq D$, $\sigma_i^{(d)}$, $i=1,\dots,N_d$, the latter being composed by $(d+1)$ nodes, 
i.e., $0$-simplices.
By choosing the connection matrices among dimensions with entries
\begin{equation}
\label{eq:Mk}
A_d(\sigma_i^{(d-1)},\sigma_j^{(d)})=1 \text{ iff } \sigma_i^{(d-1)}\subset \sigma_j^{(d)}\, ,
\end{equation}
the corresponding cross-dimensional upper and lower degrees are therefore
\begin{equation}
\label{eq:dsigmaup}
k_{i_{up}}^{(d-1)} = \sum_{j=1}^{N_d} A_d(\sigma_i^{(d-1)},\sigma_j^{(d)})\,,
\end{equation}

\begin{equation}
\label{eq:dsigmadwn}
k_{j_{down}}^{(d)} = \sum_{i=1}^{N_{d-1}} A_d(\sigma_i^{(d-1)},\sigma_j^{(d)})\equiv d+1\, ,
\end{equation}
with the total degree of simplex $\sigma_i^{(d)}$ being $k_i^{(d)} = k_{i_{up}}^{(d)} + k_{i_{down}}^{(d)}$.
With Eqs. \eqref{eq:Mk}, \eqref{eq:dsigmaup} we can so define a dynamical evolution equation for the probability as:
\begin{equation}
\label{eq:ptk}
\begin{aligned}
p_{t+1}(\sigma_i^{(d)}) = &\sum_{j=1}^{N_{d-1}} p_{t}(\sigma_j^{(d-1)})\frac{M_d(\sigma_j^{(d-1)},\sigma_i^{(d)})}{k_{j_{up}}^{(d-1)}+k_{j_{down}}^{(d-1)}}+ \\ &\sum_{l=1}^{N_{d+1}} p_{t}(\sigma_l^{(d+1)})\frac{M_{d+1}(\sigma_i^{(d)},\sigma_l^{(d+1)})}{k_{l_{up}}^{(d+1)}+k_{l_{down}}^{(d+1)}}\, .
\end{aligned}
\end{equation}
where, the first (second) term on the right-hand side disappears if $d=0$ $(d=D)$ since there are no lower (upper) dimensional simplices to jump to.

Eq. \eqref{eq:ptk} can be rewritten for convenience and compactness as

\begin{equation}
    \vec{p}_{t+1} = \vec{p}_t \,\mathbf{M}
    \label{eq:compact_rw}
\end{equation}
where the operator $\mathbf{M}$ encapsulates the normalized transitions between the various orders \cite{febbe2026random}: 
\begin{equation}
\label{eq:MatrixM}
\mathbf{M}=\left(
\begin{matrix}
 \mathbf{O}_{0} & \hat{\mathbf{M}}_1 & \dots & \dots & \dots & \dots\\
\tilde{\mathbf{M}}_1^\top & \mathbf{O}_{1} & \hat{\mathbf{M}}_2 & \dots & \dots & \dots\\
\vdots & \tilde{\mathbf{M}}_2^\top & \mathbf{O}_{2} & \hat{\mathbf{M}}_3 & \dots & \dots\\
\vdots &\vdots & \ddots &\ddots &\ddots & \dots\\
\vdots &\vdots & \vdots &\tilde{\mathbf{M}}_d^\top & \mathbf{O}_{d} & \hat{\mathbf{M}}_{d+1}\\
\vdots &\vdots & \vdots &\vdots & \ddots &\ddots\\
\end{matrix}\right)\, ,
\end{equation}

and therefore describes, via Eq. \eqref{eq:compact_rw}, a process composed of one upper-/lower-dimensional jumps occurring at discrete time steps.

\begin{figure}
    \centering
    \includegraphics[width=0.99\linewidth]{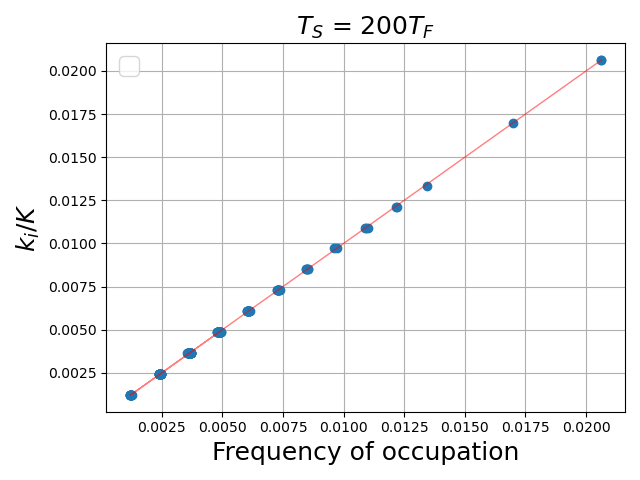}
    \caption{Correspondence between the frequency of occupation on the simplices and the corresponding (rescaled) degree by considering the simulation time $T_s >>T_F$. Here, we set the number of simulation runs to collect various random walk paths to 50.}
    \label{fig:asymptotic_state}
\end{figure}

The operator $\mathbf{M}$ evolves the random-walk state while preserving probability, with maximum eigenvalue $\lambda_{\text{max}}=1$ associated with the eigenvector whose entries are given by $k_i^{(d)}/\sum_\delta\sum_jk_j^{(\delta)}$, which also represents the asymptotic evolutionary state. This is shown in Fig. \ref{fig:asymptotic_state}, where we plot the normalized occupation frequency against the rescaled degree. For this plot, we took the simulation time $T_s \gg T_F$, where the Fiedler time $T_F$ is depends on the second-largest eigenvalue of the evolution operator.

In order to compute the degree normalization, corresponding to the total degree of all simplices across all dimensions, namely $K=\sum_{d=0}^D\sum_jk_j^{(d)}$, we can proceed iteratively by expressing the degree sum at each fixed dimension in terms of the number of structures present into the simplicial complex.

Let us pose for simplicity of exposition $D=3$, given a simplicial complex with $N_0$ nodes, $N_1$ links, $N_2$ triangles, and $N_3$ tetrahedra 
(this computation can be readily extended to a general maximum dimension), we can write the number of structures in terms of lower-degree simplices 
as (see Fig. \ref{fig:degree_to_number_structures}):
\begin{figure*}[h!!]
    \centering
    \includegraphics[width=0.9\linewidth]{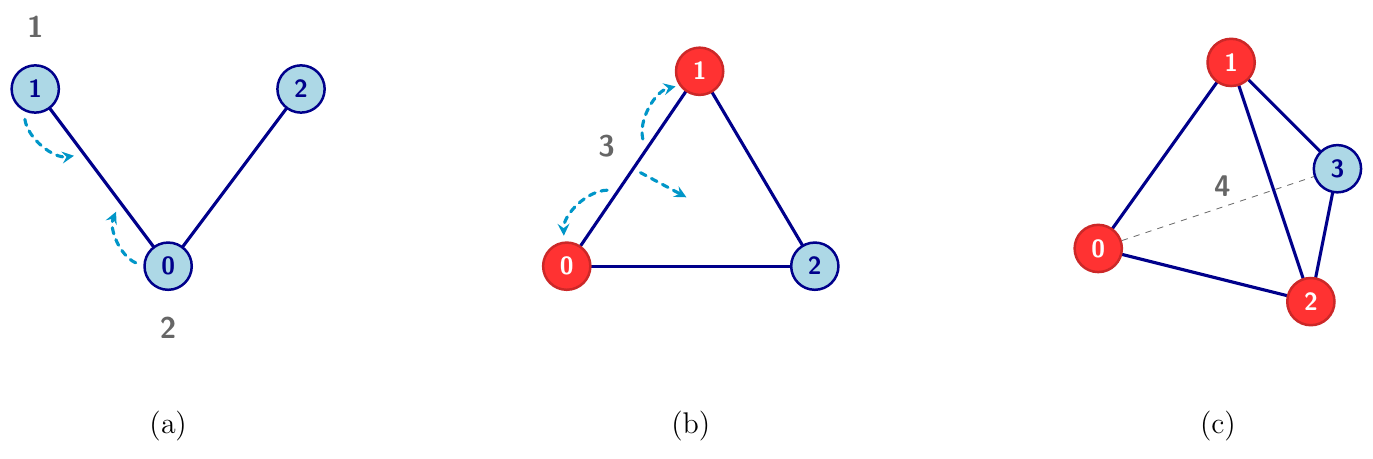}
    \caption{Here we show how the number of structures in the simplicial complex can be counted as a function of the degree of 
    lower-dimensional simplices. In panel (a), we see how the number of links is counted once for each connected node. 
    In panel (b), we see that the degree of the link $[0,1]$ can be counted as the sum of the two lower-dimensional structures 
    (nodes 0 and 1) plus the number of incident upper connected triangles. In panel (c), we see how the degree of the triangle $[0,1,2]$ 
    can be expressed as the three boundary links plus the number of incident upper connected tetrahedra. 
    From this we get to Eq. \eqref{eq:n_structures_degrees}.}
    \label{fig:degree_to_number_structures}
\end{figure*}

\begin{equation}
\label{eq:n_structures_degrees}
\begin{aligned}
  N_1     &= \sum_{j=1}^{N_0} \frac{k^{(0)}_j}{2}, \\
  N_2 &= \sum_{j=1}^{N_1} \frac{k^{(1)}_j-2}{3}, \\
  N_3    &= \sum_{j=1}^{N_2} \frac{k^{(2)}_j-3}{4},
\end{aligned}
\end{equation}

which can be rearranged as, 
\begin{equation}
    \sum_{j=1}^{N_0} k^{(0)}_j + \sum_{j=1}^{N_1} k^{(1)}_j + \sum_{j=1}^{N_2} k^{(2)}_j +
    \sum_{j=1}^{N_3} k^{(3)}_j = 4N_1 + 6N_2 + 8 N_3 \equiv K.
\end{equation}

Let us generalize this computation with the extension of Eq. \eqref{eq:n_structures_degrees} for generic $d$:
\begin{equation}
    N_{d+1} = \frac{1}{d+1}\sum_{j=1}^{N_d}k_{j_{up}}^{(d)},
\end{equation}
from which, we can hence write
\begin{equation}
\begin{alignedat}{4}
& \sum_{i=1}^{N_0} k_i^{(0)}      &=\:\:\:\:& 2 N_1 \qquad+\qquad 0, \\[2pt]
&                                &   & \vdots & \\[2pt]
& \sum_{i=1}^{N_{d-1}} k_i^{(d-1)}&=\:\:\:& (d+1) N_d\:\: +\:\: d N_{d-1}, \\[2pt]
& \sum_{i=1}^{N_d} k_i^{(d)}      &=\:\:\:& (d+2) N_{d+1} \,+\, (d+1) N_d, \\[2pt]
&                                &   & \vdots \\[2pt]
& \sum_{i=1}^{D} k_i^{(D)} 
&= \:\:\:&0 \qquad \: + \qquad \: (D+1) N_D.
\end{alignedat}
\end{equation}

Note that the first term on the right-hand side of each equation equals the second term on the right-hand side of the next equation, while the zeros in the first and last equations indicate that the lower (upper) degree of $d=0$ ($d=D$) simplices is 0.  
From that follows:
\begin{equation}
    K = \sum_{d=1}^D2(d+1)N_d.
\end{equation}

\section{Optimal Search Strategy} \label{sec:OSS_approximation}
Here, we study a generalization of the dynamics defined by the operator $\mathbf{M}$ in Eq. \eqref{eq:MatrixM}, 
where the random walk, which consists of local jumps towards connected structures, 
is modulated by a parameter $\alpha$ with an additional stochastic component consisting in a global teleportation term. 
The equation describing the dynamical evolution of this process is given by:
\begin{equation}
\label{eq:ptalpha_sparse}
\vec{p}_{t+1}=\vec{p}_t\left(\alpha \mathbf{M}+(1-\alpha) D_s\mathbf{S}\right), 
\end{equation}
where $\mathbf{S}$ is a random symmetric sparse matrix, with density 
$\delta = \sum_{i,j}S_{ij}/N^2$, and $D_s=\text{diag}(\vec{k_s})^{-1}$ where the degree of simplices due to the random connections 
is $k_s=\sum_j \mathbf{S}_{ij}$, (see \cite{di2015optimal}).

By defining the operator
\begin{equation}
\label{eq:random_walk_stoc_operator}
    \mathbf{R} =\alpha \mathbf{M}+(1-\alpha) D_s\mathbf{S},
\end{equation}
the mean first passage time can be expressed as
\begin{equation}
    T_{ij} = \sum_{k=1}^{N-1}\left(\mathbf{Z_j}^{-1}\right)_{ik},
    \label{eq:time_di_patti}
\end{equation}
where $\mathbf{Z_j} = \mathbb{I}-\mathbf{R}_j$, with $\mathbf{R}_j$ denoting the sub-matrix of the operator defined in 
Eq. \eqref{eq:random_walk_stoc_operator}, obtained by excluding the $j$-th row and column (see \cite{di2015optimal, zhang2011mean, peri2026smart}).

From Eq. \eqref{eq:time_di_patti} we can consider the node explorability $\langle T \rangle$, defined by averaging the 
times over both all starting and landing $0^\textsuperscript{th}$ $\!\!\!\!$-order simplices providing us with a scalar measure of 
how easy it is to explore the whole structure, with the parameter $\alpha$ being the reciprocal mixing between local process and global noise.

As shown by the celebrated result \cite{page1999pagerank} as well as in \cite{di2015optimal, febbe2026random} for different topologies, 
the competitive effect between these two processes can yield an optimal value for network (or simplicial complex) navigation for a 
specific value of $\alpha$. Here we present a perturbative scheme for the simplicial complex node explorability to eventually recovering this optimal search strategy configuration.

By following the computation presented in \cite{di2015optimal} we can write $\langle T \rangle$, 
as a series expansion explicitly dependent on the modulating parameter $\alpha$ by approximating the inverse matrix present in 
Eq. \eqref{eq:time_di_patti} 

\begin{equation}
    \langle T \rangle = \sum_{\ell=0} c_\ell \alpha^\ell,
    \label{eq:explorability_series_approx}
\end{equation}

with $c_\ell$ suitable coefficients depending on $\mathbf{M}$ and $\mathbf{S}$, but independent of $\alpha$ (see Appendix \ref{sec:appendix} for the mathematical derivation).

\begin{figure} [h!]
    \centering
    \includegraphics[width=0.8\linewidth]{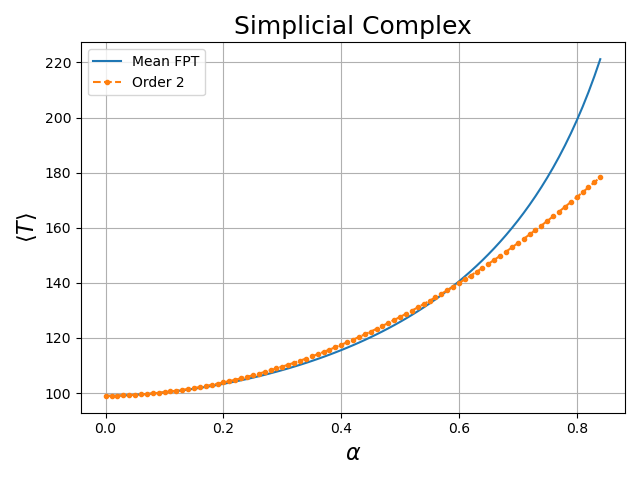}
    \caption{Node explorability $\langle T \rangle$ (solid line in the figure), computed using Eq. \eqref{eq:time_di_patti}, 
    for a simplicial complex with $N_0=50$ nodes, $D=2$, and a teleportation matrix $\mathbf{S}$ with density $\delta=1$ 
    together with the second-order approximation $(\ell \leq 2)$ of the series expansion in powers of $\alpha$ reported in 
    Eq. \eqref{eq:explorability_series_approx} (dashed orange line with filled circles).}
    \label{fig:explorability_second_order_approximation}
\end{figure}

\begin{figure*} [h!!]
    \centering
    \subfloat[]{\includegraphics[width=0.49\linewidth]{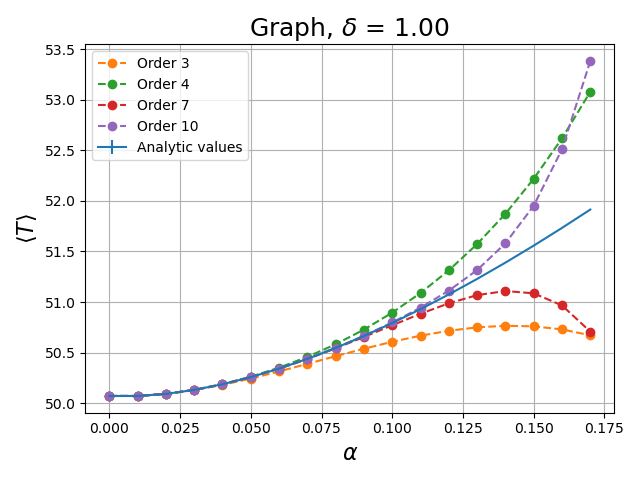}}
    \subfloat[]{\includegraphics[width=0.49\linewidth]{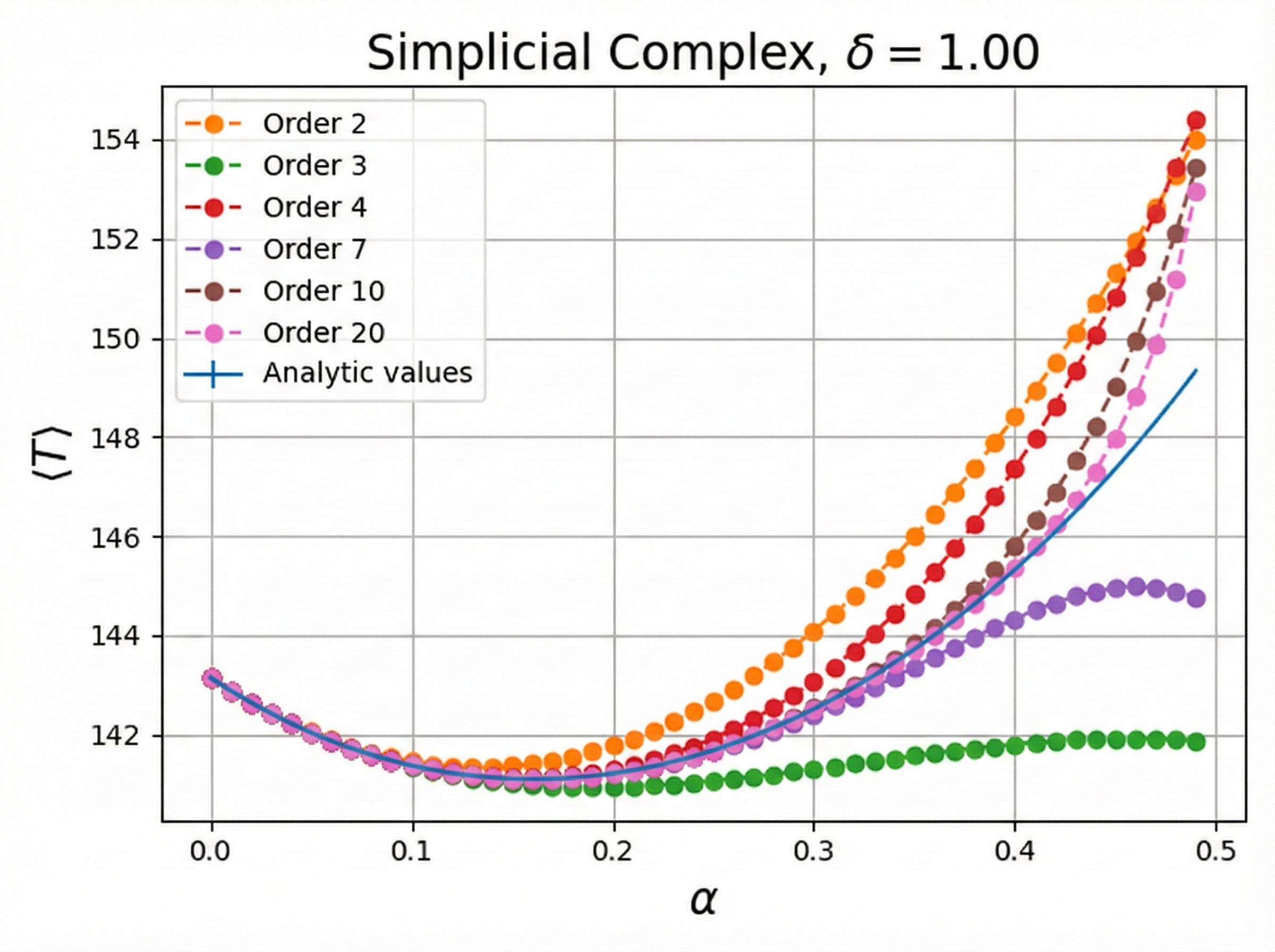}}
    \caption{Here, we report the mean node explorability $\langle T \rangle$ (solid line in figure), computed using Eq. \eqref{eq:time_di_patti}, 
    for a network (panel (a)) and a simplicial complex (panel (b)) with $N_0=20$ nodes, construction parameter $p_3=1$ (see \cite{febbe2026model}),
     and a teleportation matrix $\mathbf{S}$ with density $\delta=1$. 
     As we can see, successive approximations of the series in $\alpha$ (according to Eq. \ref{eq:appendix_alpha_serie}) 
     increasingly match the explorability curves, eventually reproducing the minima (see panel (b)). 
     Interestingly, we observe a clear contrast in behavior between the network and simplicial complex topologies in the regime $\alpha \to 0$: 
     the explorability increases in the former, while it decreases in the latter as the contribution of the local random walk increases.}
    \label{fig:explorability_superior_order_approximation}
\end{figure*}

By truncating the series reported in Eq. \eqref{eq:explorability_series_approx} at $\ell\leq2$, we obtain the expansion up to the second order.
An example of this result is shown in Fig. \ref{fig:explorability_second_order_approximation}, while 
Fig. \ref{fig:explorability_superior_order_approximation} displays various successive approximation terms to extend the range in $\alpha$ over which the expansion proves adequate. Interestingly enough, by progressively including higher order terms, the series approximation yields the correct location of the minimum for $\langle T\rangle$ vs. alpha, i.e. the optimal strategy value for  the simplicial complex (panel (b)) settings. Notice, in fact, that a sharp differences emerges from the inspection of Fig. \ref{fig:explorability_superior_order_approximation}, when comparing network and simplicial complex settings (the former being the skeleton of the latter, is the structure after sweeping out the higher order interactions). 
For the binary topology, the minimum of $\langle T \rangle$ occurs, 
in this case, at $\alpha = 0$ (pure global stochasticity), implying that the exploration time increases as soon as local dynamics are introduced. 
Conversely, for the simplicial complex, the minimum shifts to a value well within the interval $(0,1)$ implying a decreasing trend 
for small values of $\alpha$, materializing in an optimized search strategy characterized by a non-trivial mix of the two fundamental dynamical processes considered and combined via 
Eq. \eqref{eq:random_walk_stoc_operator}. 
This scenario suggests a qualitative difference in behavior between the two imposed topologies analyzed in Fig. \ref{fig:explorability_superior_order_approximation}. 
Consequently, specific dynamical responses, like those presented throughout this section and in \cite{febbe2026random}, 
could serve, in specific contexts, as a diagnostic signature to distinguish the underlying topology given the observed dynamics.

\section{\label{sec:Conclusion} Conclusions}
In recent years, higher-order structures have become increasingly popular in the modeling of complex interacting systems, and the analysis of dynamical processes interacting with their topology is of timely interest.
Here we discuss and delve deeply into the explorability of simplicial complexes via cross-dimensional random dynamics.
The random walk process, proposed in \cite{febbe2026random}, has been analyzed in terms of the mean first passage time among the nodes, with the addition of a global stochastic teleportation term modulated via a parameter $\alpha$.
The ensuing optimal search combination (i.e. the one that minimize $\langle T \rangle $ vs. $\alpha$) can be iteratively identified by using a properly calibrated approximation scheme.
Interestingly, an instance of different behaviour is highlighted by comparing the response of simplicial complexes to dynamics with that of their binary skeletons. This provides yet another example of the fundamental interplay between dynamical processes and the underlying topology on which they are bound to.

\section*{Acknowledgments}    
The authors would like to acknowledge the support by \textsc{\#nextgenerationeu} (\textsc{ngeu}) funded by the Ministry of University and Research (\textsc{mur}), National Recovery and Resilience Plan (\textsc{nrrp}), project \textsc{mnesys} (PE0000006)—A multiscale integrated approach to the study of the nervous system in health and disease (DN. 1553 11.10.2022).

\appendices
\section{Perturbative scheme for exploration time}\label{sec:appendix}
Here we adapt the computation carried out in \cite{di2015optimal} in order to obtain an expansion of the node explorability $\langle T \rangle$ 
for simplicial complexes in powers of the modulating parameter $\alpha$.

Let us consider $C$ and $B$ to be two arbitrary non-singular square matrices of dimension $N$. Let us start by noting that the following identity 
holds:
\begin{equation}
(C + \varepsilon B)^{-1} = \left( \mathbb{I}_n + \varepsilon C^{-1} B \right)^{-1} C^{-1},
\end{equation}
and expressing $\mathbb{I}_n + \varepsilon C^{-1} B$ as a Neumann series \cite{stewart1998matrix}, it follows
\begin{equation}
\begin{aligned}
(C + \varepsilon B)^{-1}
= &C^{-1}
- \varepsilon C^{-1} B C^{-1}
+
\varepsilon^2 C^{-1} B C^{-1} B C^{-1}
\\
& +\ldots +\varepsilon^n C^{-1} (B C^{-1})^n+ \dots
\end{aligned}
\end{equation}
by truncating it at the proper order, it yields the desired approximation.

Consequently, by denoting by $j$ the position of the trap on the nodes, the terms of the reduced matrix $\mathbf{Z}_j$ 
can be straightforwardly rearranged by grouping together those proportional to $\alpha$. Explicitly, one obtains:
\begin{equation}
Z_j = \mathbb{I}_{N-1} - (D_s^{-1})_j S_j
+ \alpha \left[ (D_s^{-1})_j S_j - M_j^{-1} \right].
\end{equation}

Setting
\[
C_j = \mathbb{I}_{N-1} - (D_s^{-1})_j S_j,
\qquad
B_j = (D_s^{-1})_j S_j - M_j^{-1},
\]
and $\varepsilon = \alpha$, we can write the general coefficient of the expansion
\begin{equation}
    \langle T \rangle = \sum_{\ell=0} c_\ell \alpha^\ell
    \label{eq:appendix_alpha_serie}
\end{equation}
that can be truncated at the desired order,

\begin{equation}
    c_\ell = \frac{1}{N_0(N_0-1)}\sum_{1\leq i\ne j \leq N_0} \sum_k \left( C_j^{-1} \left( B_j C_j^{-1} \right)^\ell \right)_{ik}.
\end{equation}

Fig. \ref{fig:explorability_superior_order_approximation} shows different-order approximations of the explorability curves. 
Including more terms, we can follow the behaviour for larger values of the modulating parameter $\alpha$ and better reproduce the 
minima corresponding to the optimal search strategy.



\bibliographystyle{IEEEtran}
\bibliography{bibliography}

@article{vasilyeva2021multilayer,
  title={Multilayer representation of collaboration networks with higher-order interactions},
  author={Vasilyeva, E and Kozlov, A and Alfaro-Bittner, Karin and Musatov, D and Raigorodskii, AM and Perc, Matja{\v{z}} and Boccaletti, Stefano},
  journal={Scientific reports},
  volume={11},
  number={1},
  pages={5666},
  year={2021},
  publisher={Nature Publishing Group UK London}
}

@inproceedings{fujisawa2022analyzing,
  title={Analyzing configuration transitions associated with higher-order link occurrences in networks of cooking ingredients},
  author={Fujisawa, Koudai and Kumano, Masahito and Kimura, Masahiro},
  booktitle={International Conference on Complex Networks and Their Applications},
  pages={623--635},
  year={2022},
  organization={Springer}
}

@book{newman2010networks,
 title={Networks: An introduction},
 author={Newman, Mark EJ},
 year={2010},
 publisher={Oxford university press}
}

@book{barabasi2016network,
  title={Network Science},
  author={Barab{\'a}si, Albert-L{\'a}szl{\'o}},
  publisher={Cambridge University Press},
  year={2016},
  url={http://networksciencebook.com/}
}

@article{zhang2005general,
  title={A general framework for weighted gene co-expression network analysis},
  author={Zhang, Bin and Horvath, Steve and others},
  journal={Statistical applications in genetics and molecular biology},
  volume={4},
  number={1},
  pages={1128},
  year={2005}
}

@article{barabasi2023neuroscience,
  title={Neuroscience needs network science},
  author={Barab{\'a}si, D{\'a}niel L and Bianconi, Ginestra and Bullmore, Ed and Burgess, Mark and Chung, SueYeon and Eliassi-Rad, Tina and George, Dileep and Kov{\'a}cs, Istv{\'a}n A and Makse, Hern{\'a}n and Nichols, Thomas E and others},
  journal={Journal of Neuroscience},
  volume={43},
  number={34},
  pages={5989--5995},
  year={2023},
  publisher={Society for Neuroscience}
}

@article{lucas2023inferring,
  title={Inferring cell cycle phases from a partially temporal network of protein interactions},
  author={Lucas, Maxime and Morris, Arthur and Townsend-Teague, Alex and Tichit, Laurent and Habermann, Bianca and Barrat, Alain},
  journal={Cell Reports Methods},
  volume={3},
  number={2},
  year={2023},
  publisher={Elsevier}
}

@inproceedings{febbe2024chaos,
  title={Chaos and Synchronization in the UJT Relaxation Oscillator},
  author={Febbe, Diego and Di Garbo, Angelo and Mannella, Riccardo and Meucci, Riccardo and Fanelli, Duccio},
  booktitle={2024 IEEE Workshop on Complexity in Engineering (COMPENG)},
  pages={1--7},
  year={2024},
  organization={IEEE}
}

@inproceedings{buongiovanni2022will,
  title={Will you take the knee? italian twitter echo chambers’ genesis during euro 2020},
  author={Buongiovanni, Chiara and Candusso, Roswita and Cerretini, Giacomo and Febbe, Diego and Morini, Virginia and Rossetti, Giulio},
  booktitle={International Conference on Complex Networks and Their Applications},
  pages={29--40},
  year={2022},
  organization={Springer}
}

@article{nishikawa2015comparative,
  title={Comparative analysis of existing models for power-grid synchronization},
  author={Nishikawa, Takashi and Motter, Adilson E},
  journal={New Journal of Physics},
  volume={17},
  number={1},
  pages={015012},
  year={2015},
  publisher={IOP Publishing}
}

@article{buffoni2022spectral,
  title={Spectral pruning of fully connected layers},
  author={Buffoni, Lorenzo and Civitelli, Enrico and Giambagli, Lorenzo and Chicchi, Lorenzo and Fanelli, Duccio},
  journal={Scientific Reports},
  volume={12},
  number={1},
  pages={11201},
  year={2022},
  publisher={Nature Publishing Group UK London}
}

@article{chicchi2026estimating,
  title={Estimating global input relevance and enforcing sparse representations with a scalable spectral neural network approach},
  author={Chicchi, Lorenzo and Buffoni, Lorenzo and Febbe, Diego and Giambagli, Lorenzo and Marino, Raffaele and Fanelli, Duccio},
  journal={Communications Physics},
  year={2026},
  publisher={Nature Publishing Group UK London}
}

@article{corso2024graph,
  title={Graph neural networks},
  author={Corso, Gabriele and Stark, Hannes and Jegelka, Stefanie and Jaakkola, Tommi and Barzilay, Regina},
  journal={Nature Reviews Methods Primers},
  volume={4},
  number={1},
  pages={17},
  year={2024},
  publisher={Nature Publishing Group UK London}
}

@article{centola2010spread,
  title={The spread of behavior in an online social network experiment},
  author={Centola, Damon},
  journal={science},
  volume={329},
  number={5996},
  pages={1194--1197},
  year={2010},
  publisher={American Association for the Advancement of Science}
}

@article{pacelli2025link,
  title={The link between climate and systemic risk: A bibliometric and systematic literature review},
  author={Pacelli, Vincenzo and Foglia, Matteo and Mariano, Dayana},
  journal={Research in International Business and Finance},
  pages={103072},
  year={2025},
  publisher={Elsevier}
}

@article{lucarini2026synergistic,
  title={The Synergistic Route to Stretched Criticality},
  author={Lucarini, Lorenzo and Meloni, Sandro and Villegas, Pablo},
  journal={arXiv preprint arXiv:2604.28003},
  year={2026}
}

@article{bianconi2025gravity,
  title={Gravity from entropy},
  author={Bianconi, Ginestra},
  journal={Physical Review D},
  volume={111},
  number={6},
  pages={066001},
  year={2025},
  publisher={APS}
}

@article{iacopini2019simplicial,
  title={Simplicial models of social contagion},
  author={Iacopini, Iacopo and Petri, Giovanni and Barrat, Alain and Latora, Vito},
  journal={Nature communications},
  volume={10},
  number={1},
  pages={2485},
  year={2019},
  publisher={Nature Publishing Group UK London}
}

@inproceedings{morris2019weisfeiler,
  title={Weisfeiler and leman go neural: Higher-order graph neural networks},
  author={Morris, Christopher and Ritzert, Martin and Fey, Matthias and Hamilton, William L and Lenssen, Jan Eric and Rattan, Gaurav and Grohe, Martin},
  booktitle={Proceedings of the AAAI conference on artificial intelligence},
  volume={33},
  number={01},
  pages={4602--4609},
  year={2019}
}

@article{peri2026spectral,
  title={Spectral Higher-Order Neural Networks},
  author={Peri, Gianluca and Carletti, Timoteo and Fanelli, Duccio and Febbe, Diego},
  journal={arXiv preprint arXiv:2603.28420},
  year={2026}
}

@article{giusti2016two,
  title={Two’s company, three (or more) is a simplex: Algebraic-topological tools for understanding higher-order structure in neural data},
  author={Giusti, Chad and Ghrist, Robert and Bassett, Danielle S},
  journal={Journal of computational neuroscience},
  volume={41},
  number={1},
  pages={1--14},
  year={2016},
  publisher={Springer}
}

@article{battiston2020networks,
  title={Networks beyond pairwise interactions: Structure and dynamics},
  author={Battiston, Federico and Cencetti, Giulia and Iacopini, Iacopo and Latora, Vito and Lucas, Maxime and Patania, Alice and Young, Jean-Gabriel and Petri, Giovanni},
  journal={Physics reports},
  volume={874},
  pages={1--92},
  year={2020},
  publisher={Elsevier}
}

@article{carletti2023global,
  author  = {Carletti, Timoteo and Giambagli, Lorenzo and Bianconi, Ginestra},
  title   = {Global Topological Synchronization on Simplicial and Cell Complexes},
  journal = {Physical Review Letters},
  volume  = {130},
  pages   = {187401},
  year    = {2023}
}

@article{gambuzza2021stability,
  author  = {Gambuzza, Lucia Valentina and Di Patti, Francesco and Gallo, Luca and Lepri, Stefano and Romance, Miguel and Criado, Regino and Frasca, Mattia and Latora, Vito and Boccaletti, Stefano},
  title   = {Stability of synchronization in simplicial complexes},
  journal = {Nature Communications},
  volume  = {12},
  pages   = {1255},
  year    = {2021}
}

@article{muolo2023turing,
  title={Turing patterns in systems with high-order interactions},
  author={Muolo, Riccardo and Gallo, Luca and Latora, Vito and Frasca, Mattia and Carletti, Timoteo},
  journal={Chaos, Solitons \& Fractals},
  volume={166},
  pages={112912},
  year={2023},
  publisher={Elsevier}
}

@article{giambagli2022diffusion,
  title={Diffusion-driven instability of topological signals coupled by the Dirac operator},
  author={Giambagli, Lorenzo and Calmon, Lucille and Muolo, Riccardo and Carletti, Timoteo and Bianconi, Ginestra},
  journal={Physical Review E},
  volume={106},
  number={6},
  pages={064314},
  year={2022},
  publisher={APS}
}

@inproceedings{muolo2024turing,
  title={Turing patterns on discrete topologies: from networks to higher-order structures},
  author={Muolo, Riccardo and Giambagli, Lorenzo and Nakao, Hiroya and Fanelli, Duccio and Carletti, Timoteo},
  booktitle={Proceedings A},
  volume={480},
  number={2302},
  pages={20240235},
  year={2024},
  organization={The Royal Society}
}

@article{moriame2025hamiltonian,
  title={Hamiltonian control to desynchronize Kuramoto oscillators with higher-order interactions},
  author={Moriam{\'e}, Martin and Lucas, Maxime and Carletti, Timoteo},
  journal={Physical Review E},
  volume={111},
  number={4},
  pages={044307},
  year={2025},
  publisher={APS}
}

@article{millan2025topology,
  title={Topology shapes dynamics of higher-order networks},
  author={Mill{\'a}n, Ana P and Sun, Hanlin and Giambagli, Lorenzo and Muolo, Riccardo and Carletti, Timoteo and Torres, Joaqu{\'\i}n J and Radicchi, Filippo and Kurths, J{\"u}rgen and Bianconi, Ginestra},
  journal={Nature Physics},
  volume={21},
  number={3},
  pages={353--361},
  year={2025},
  publisher={Nature Publishing Group UK London}
}

@article{gallo2022synchronization,
  title={Synchronization induced by directed higher-order interactions},
  author={Gallo, Luca and Muolo, Riccardo and Gambuzza, Lucia Valentina and Latora, Vito and Frasca, Mattia and Carletti, Timoteo},
  journal={Communications Physics},
  volume={5},
  number={1},
  pages={263},
  year={2022},
  publisher={Nature Publishing Group UK London}
}

@book{bianconi2021higher,
  title={Higher-order networks},
  author={Bianconi, Ginestra},
  year={2021},
  publisher={Cambridge University Press}
}

@article{leon2024higher,
  title={Higher-order interactions induce anomalous transitions to synchrony},
  author={Le{\'o}n, Iv{\'a}n and Muolo, Riccardo and Hata, Shigefumi and Nakao, Hiroya},
  journal={Chaos: An Interdisciplinary Journal of Nonlinear Science},
  volume={34},
  number={1},
  year={2024},
  publisher={AIP Publishing}
}

@article{bianconi2016network,
  title={Network geometry with flavor: from complexity to quantum geometry},
  author={Bianconi, Ginestra and Rahmede, Christoph},
  journal={Physical Review E},
  volume={93},
  number={3},
  pages={032315},
  year={2016},
  publisher={APS}
}

@article{bianconi2017emergent,
  title={Emergent hyperbolic network geometry},
  author={Bianconi, Ginestra and Rahmede, Christoph},
  journal={Scientific reports},
  volume={7},
  number={1},
  pages={41974},
  year={2017},
  publisher={Nature Publishing Group UK London}
}

@article{torres2020simplicial,
  title={Simplicial complexes: higher-order spectral dimension and dynamics},
  author={Torres, Joaqu{\'\i}n J and Bianconi, Ginestra},
  journal={Journal of Physics: Complexity},
  volume={1},
  number={1},
  pages={015002},
  year={2020},
  publisher={IOP Publishing}
}

@article{febbe2026model,
  title={Model of Simplicial Complexes with dimension-wise preferential attachment},
  author={Febbe, Diego and Fanelli, Duccio and Carletti, Timoteo},
  journal={arXiv preprint arXiv:2605.17004},
  year={2026}
}

@article{febbe2026random,
  title={Random Walks Across Dimensions: Exploring Simplicial Complexes},
  author={Febbe, Diego and Fanelli, Duccio and Carletti, Timoteo},
  journal={arXiv preprint arXiv:2601.16086},
  year={2026}
}

@article{petri2014homological,
  title={Homological scaffolds of brain functional networks},
  author={Petri, Giovanni and Expert, Paul and Turkheimer, Federico and Carhart-Harris, Robin and Nutt, David and Hellyer, Peter J and Vaccarino, Francesco},
  journal={Journal of The Royal Society Interface},
  volume={11},
  number={101},
  pages={20140873},
  year={2014}
}

@techreport{page1999pagerank,
  title={The PageRank Citation Ranking: Bringing Order to the Web},
  author={Page, Lawrence and Brin, Sergey and Motwani, Rajeev and Winograd, Terry},
  year={1999},
  institution={Stanford InfoLab},
  number={1999-66},
  url={http://ilpubs.stanford.edu:8090/422/}
}

@article{di2015optimal,
  title={Optimal search strategies on complex multi-linked networks},
  author={Di Patti, Francesca and Fanelli, Duccio and Piazza, Francesco},
  journal={Scientific reports},
  volume={5},
  number={1},
  pages={9869},
  year={2015},
  publisher={Nature Publishing Group UK London}
}

@article{zhang2011mean,
  title={Mean first-passage time for random walks on undirected networks},
  author={Zhang, Zhongzhi and Julaiti, Alafate and Hou, Baoyu and Zhang, Hongjuan and Chen, Guanrong},
  journal={The European Physical Journal B},
  volume={84},
  number={4},
  pages={691--697},
  year={2011},
  publisher={Springer}
}

@article{peri2026smart,
title = {Smart walkers in discrete space},
journal = {Chaos, Solitons \& Fractals},
volume = {208},
pages = {118356},
year = {2026},
issn = {0960-0779},
author = {Gianluca Peri and Lorenzo Buffoni and Giacomo Chiti and Duccio Fanelli and Raffaele Marino and Andrea Nocentini and Pier Paolo Panti},
}

@article{bianconi2021topological,
  title={The topological Dirac equation of networks and simplicial complexes},
  author={Bianconi, Ginestra},
  journal={Journal of Physics: Complexity},
  volume={2},
  number={3},
  pages={035022},
  year={2021},
  publisher={IOP Publishing}
}

@book{stewart1998matrix,
  title={Matrix Algorithms: Volume 1: Basic Decompositions},
  author={Stewart, G. W.},
  year={1998},
  publisher={SIAM},
  address={Philadelphia, PA},
  series={Other Titles in Applied Mathematics}
}

@article{munoz2025exploring,
  title={Exploring the limits of the law of mass action in the mean field description of epidemics on Erd{\"o}s-R{\'e}nyi networks},
  author={Mu{\~n}oz, Francisco J and Meacci, Luca and Nu{\~n}o, Juan Carlos and Primicerio, Mario},
  journal={Applied Mathematics and Computation},
  volume={485},
  pages={129019},
  year={2025},
  publisher={Elsevier}
}

@article{Benson-2018-simplicial,
 author = {Benson, Austin R. and Abebe, Rediet and Schaub, Michael T. and Jadbabaie, Ali and Kleinberg, Jon},
 title = {Simplicial closure and higher-order link prediction},
 year = {2018},
 doi = {10.1073/pnas.1800683115},
 publisher = {National Academy of Sciences},
 issn = {0027-8424},
 journal = {Proceedings of the National Academy of Sciences}
}

@misc{peri2026spectralhigherorderneuralnetworks,
      title={Spectral Higher-Order Neural Networks Have Sharp Expressivity Bounds}, 
      author={Gianluca Peri and Diego Febbe and Duccio Fanelli},
      year={2026},
      eprint={2607.19042},
      archivePrefix={arXiv},
      primaryClass={cs.LG},
      url={https://arxiv.org/abs/2607.19042}, 
}

@software{diegofebbe_2026_21563618,
  author       = {Diego Febbe},
  title        = {diegofebbe/Random\_walk\_on\_simplicial\_complexes:
                   Random walk on simplicial complexes 1.0
                  },
  year         = 2026,
  publisher    = {Zenodo},
  version      = {v1.0},
  note = {10.5281/zenodo.21563618},
  doi          = {10.5281/zenodo.21563618},
}
%

\end{document}